**Influence of refraction on wind turbine noise**

Rufin Makarewicz

Institute of Acoustics

A.Mickiewicz University

61-614 Poznan, Umultowska 85,

Poland

e-mail: makaaku@amu.edu.pl

**ABSTRACT**

A semi-empirical method is applied to calculate the time-average sound level of wind turbine noise generation and propagation. Both are affected by wind shear refraction. Under upwind conditions the partially ensonified zone separates the fully ensonified zone (close to the turbine) and the shadow zone (far away from the turbine). Refraction is described in terms of the wind speed linear profile fitted to the power law profile. The rotating blades are treated as a two-dimensional circular source in the vertical plane. Inside the partially ensonified zone the effective A-weighted sound power decreases to zero when the receiver moves from the turbine toward the shadow zone. The presented results would be useful in practical applications to give a quick estimate of the effect of refraction on wind turbine noise.

## 1. INTRODUCTION

A large increase in wind power is mandated all over the world. However, this is a fast growing problem across the world, because wind turbine noise is more annoying than road traffic noise at the same level. In many countries the noise limits of the wind turbine noise are expressed in terms of the time average sound level, $L_{AeqT}$ [1].

Unfortunately, even far away turbine noise evokes annoyance, so refraction can't be ignored. Figs.1 and 2 illustrate the influence of wind induced refraction on noise propagation (Sect.2).

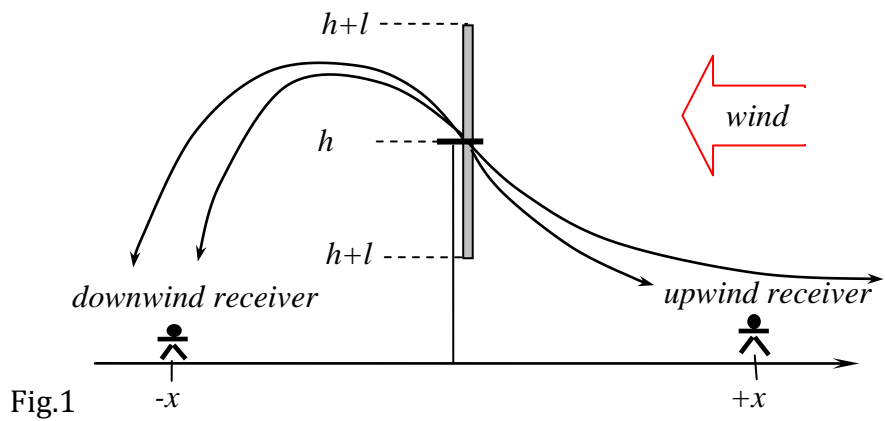

Fig.1

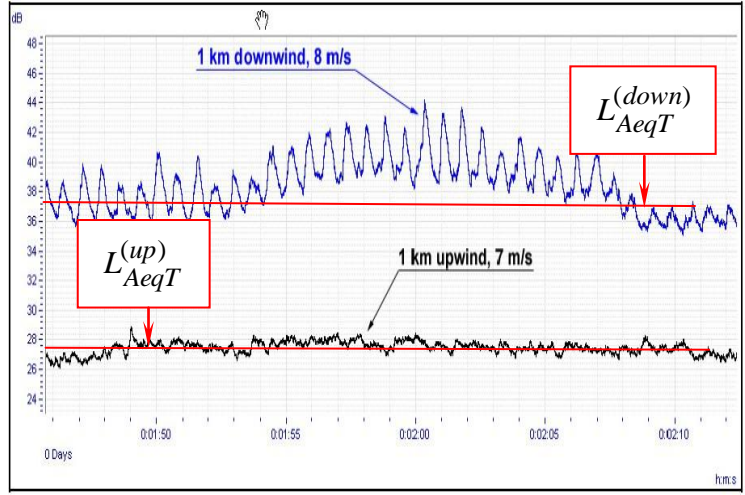

Fig.2

As expected, for downwind propagation the value of $L_{AeqT}^{(down)}$ exceeds the upwind value of $L_{AeqT}^{(up)}$ at the same horizontal distance from the turbine. The purpose of this study is to find the dependence of $L_{AeqT}$ on the horizontal distance $\rho$. The horizontal distance $\rho$ to the shadow zone (without turbine noise inside) depends, among other things, on the source height $z$ [3,4]. In reality the wind turbine sound is emitted by a rotor plane of radius $l$ and the heights of its strips vary between $h-l$ and $h+l$ (Fig.3). Consequently, the rays from the low- and high tips of the plane correspond to the shadow boundaries at horizontal distances $\rho_1$ and $\rho_2$.

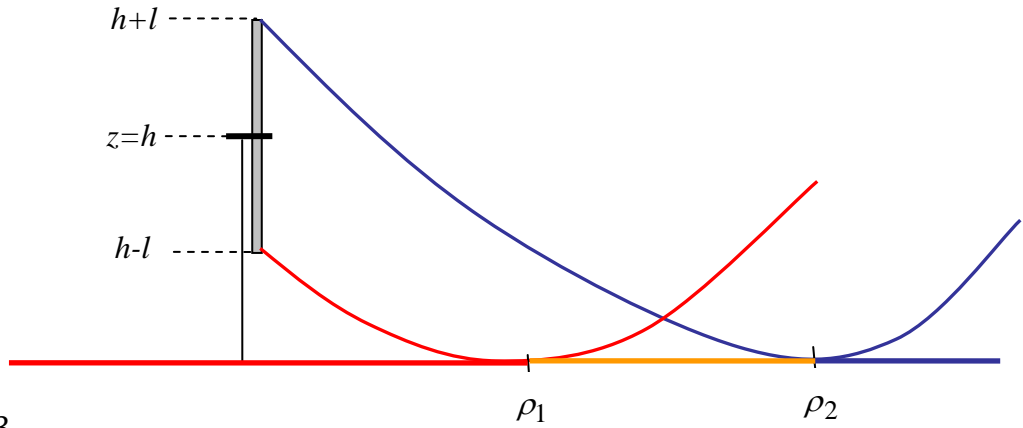

Fig.3

Finally, inequalities,

$$\rho_1 < \rho < \rho_2, \qquad (1)$$

determine the transition zone between the fully ensonified zone, $\rho < \rho_1$ (red line), and the shadow zone, $\rho > \rho_2$ (blue line). Within the partially insonified zone the time average sound level, $L_{AeqT}(\rho)$ decreases from $L_{AeqT}(\rho_1)$ at the horizontal distance $\rho_1$ to $L_{AeqT}(\rho_2) = -\infty$ dB (silence) at the distance $\rho_2$. In other words, the receiver motion from $\rho_1$ to $\rho_2$ is accompanied by the rotor plane "shrinking" from $z = h - l$ to $z = h + l$ (Sect.5). The ideas presented here come from Ref. [5].

The generation of wind turbine noise has been investigated in both experimental and theoretical studies (see Ref. [6] and the literature cited therein). There are empirical relationships between the A-weighted sound power level and the blade tip speed, rotational speed, and hub wind speed. The semi-empirical model presented in Sect. 3 employs these relationships.

## 2. REFRACTION

Refraction is quite a well-known phenomenon. The limiting rays graze the ground and make a shadow boundary in the vertical plane (Fig.4). To determine distance to the shadow zone, we apply the effective sound speed in a stratified atmosphere [3,7,8],

$$c(z) = c_{ad}(z) - V(z)\cos\alpha. \qquad (2)$$

Here $c_{ad}$ is the adiabatic speed which depends on the air temperature and the product $V(z)\cos\alpha$ denotes the wind speed component in the direction of sound propagation from the source to receiver (Fig.4).

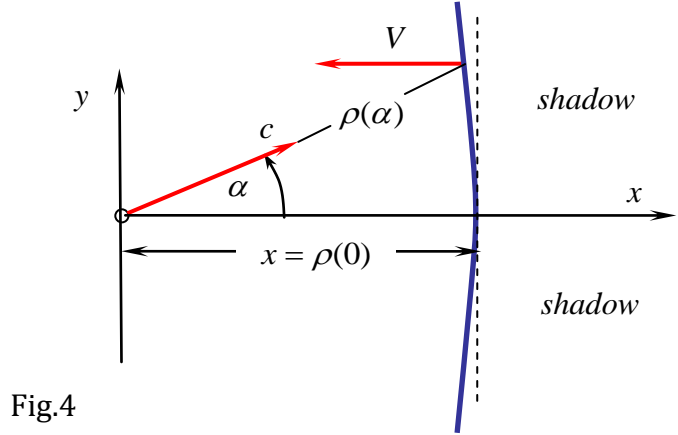

Fig.4

For wind turbine engineering the variation in wind speed with the height $z$ can be described by

$$V(z) = V(h) \cdot \left(\frac{z}{h}\right)^{\beta}, \qquad (3)$$

where $V(h)$ denotes the wind speed at the hub height $h$ (Fig.3). Two wind speed measurements at two heights, $V(z_1)$ and $V(z_2)$, yield the wind shear exponent $\beta$ (Eq.3). This exponent is a function of surface roughness and air stability. In Ref. [9] one can find that for urban and rural areas $0.15 < \beta < 0.3$ and $0.07 < \beta < 0.55$, respectively.

To obtain the final equations as simply as possible, we fit a linear profile

$$\tilde{V}(z) = V(h) \cdot \left[a\frac{z}{h} + b\right], \qquad (4)$$

to the power law wind profile (Eq.3). The unknown parameters $a$ and $b$ minimize the average error,

$$\sigma^2(a,b) = \frac{1}{h}\int_0^h \left[V(z) - \tilde{V}(z)\right]^2 dz, \qquad (5)$$

and the conditions,

$$\frac{\partial \sigma^2}{\partial a} = 0 \quad \text{and} \quad \frac{\partial \sigma^2}{\partial b} = 0, \qquad (6)$$

lead to

$$a = \frac{6\beta}{(1+\beta)(2+\beta)}, \qquad b = \frac{2(1-\beta)}{(1+\beta)(2+\beta)}. \qquad (7)$$

In this study we assume that the adiabatic component of the vertical gradient, $dc_{ad}/dz$, is relatively small as compared with the vertical gradient of the wind speed component (Eq.4),

$$\frac{dV}{dz} \approx a \cdot \frac{V(h)}{h}. \qquad (8)$$

Consequently, the linear approximation of the effective sound speed in a stratified atmosphere (Eq.2) takes the form,

$$c(z) \approx c_{ad}(0) - \frac{6\beta}{(1+\beta)(2+\beta)} \frac{V(h)\cos\alpha}{h} \cdot z. \qquad (9)$$

Here $c_{ad}(0)$ expresses the adiabatic speed at ground level (e.g. $c_{ad}(0) = 340$ m/s for the air temperature of $15\,^{o}$ C). Finally, for the point source at the height $z$, the horizontal distance to the receiver $\rho$ (Fig.4) can be calculated from [3,4],

$$\rho(\alpha) \approx \mu \sqrt{\frac{zh}{\cos\alpha}}, \qquad (10)$$

where

$$\mu = \sqrt{\frac{(1+\beta)(2+\beta)}{3\beta} \frac{c_{ad}(0)}{V(h)}}. \qquad (11)$$

If the receiver moves along the x axis then $\alpha = 0$ and $\rho(0) = x$ (Fig.4). Accordingly, the location of the partially insonified zone is determined by (Fig.3, Eq.10),

$$x_1 = \rho_1 \approx \mu\sqrt{h\cdot(h-l)} \quad \text{and} \quad x_2 = \rho_2 \approx \mu\sqrt{h\cdot(h+l)}. \qquad (12)$$

**Example 1**

For weakly stable air conditions in an urban area, $\beta = 0.3$. Then the adiabatic speed close to the ground surface $a_{ad}(0) = 340$ m/s and the hub height wind speed $V(h) = 15$ m/s give $\mu = 8.67$. Finally, when the turbine height $h = 100m$ and blade length $l = 40m$, then the transition zone is determined by $\rho_1 = 670m$ and $\rho_2 = 1025m$.

**3. NOISE GENERATION**

There are two categories of wind turbine sound: airfoil self-noise and inflow-turbulence noise [6]. The former is produced by blade rotation in motionless air. Accordingly the measured A-weighted sound power level can be expressed by the blade tip speed, $v(l)$ (Fig.5),

$$L_{WA} = 10m \cdot \log \frac{v(l)}{v_o} + A. \qquad v_o = 1 m/s. \qquad (13)$$

For the blade length of $l$ [m] and rotational speed $N$ [rps],

$$v(l) = 2\pi N \cdot l, \qquad (14)$$

where $N$ [rps] denotes rotational speed. So the empirical formula (13) can be rewritten as

$$L_{WA} = 10m \cdot \log \frac{N}{N_o} + A, \qquad N_o = 1 rps. \qquad (15)$$

In Refs.[10-12] one can find the measurement results that are characterized by $4 < m < 7$.

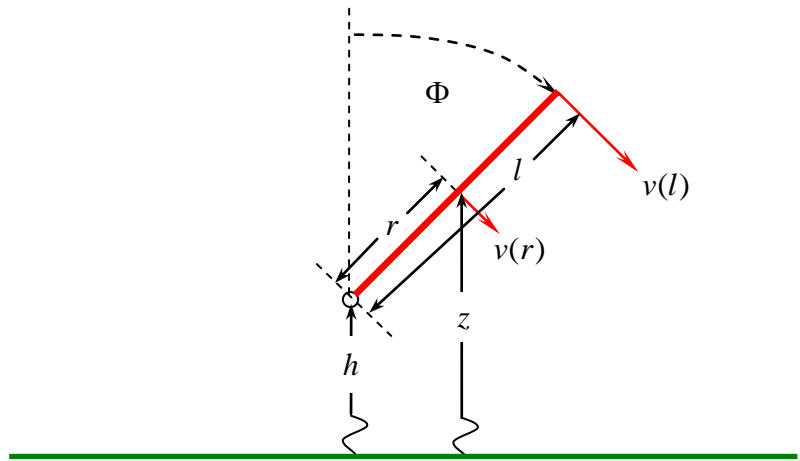

Fig.5

The definition of the A-weighted sound power level is

$$L_{WA} = 10 \log \frac{W_A}{W_o}, \qquad (16)$$

where: $W_A$ - A-weighted sound power and $W_o = 10^{-12}$ [W]. Ultimately, the empirical results given by Eq.(13) and (15) lead to the conclusion that the A-weighted sound power from a blade segment of unit length, $l_o = 1m$, at the distance $0 < r < l$ from the hub (Fig.5), is proportional to the $m$-th power of its speed $v(r)$,

$$w \propto \left(\frac{v(r)}{v_o}\right)^m. \tag{17}$$

The relationship of $v(r)$ with the blade tip speed $v(l)$,

$$v(r) = \frac{r}{l}v(l), \tag{18}$$

transforms formula (17) into,

$$w \propto \left(\frac{v(l)}{v_o}\right)^m \cdot \left(\frac{r}{l}\right)^m. \tag{19}$$

Interaction of the blade with upstream wind turbulence accounts for the second category of wind turbine sound [6]. Measurements [13-15] show that the A-weighted sound power level of turbine noise and the wind speed at the hub height, $V(h)$, are related to each other by,

$$L_{WA} = 10n \cdot \log \frac{V(h)}{v_o} + B, \quad v_o = 1m/s, \tag{20}$$

where $2 < n < 5$. The blade segment of unit length, $l_o = 1m$, sweeps through a wind that changes with the height $z$ above ground and the A-weighted sound power is proportional to the $n$-th power of the wind speed at the momentary height, $V(z)$,

$$w \propto \left(\frac{V(z)}{v_o}\right)^n. \tag{21}$$

With $z - h = r\cos\Phi$ (Fig.5), Eqs.(3) and (21) yield

$$w \propto \left(\frac{V(h)}{v_o}\right)^n \cdot F(t), \tag{22}$$

where

$$F(t) = \left[1 + \frac{r}{h}\cos\Phi\right]^{n\beta}, \quad 0 < r < l, \tag{23}$$

quantifies the wind shear effect on sound generation. Taking into account that fact that the hub height exceeds the blade length, $l << h$, the following approximation seems to be plausible:

$$F(t) = \left[1 + \frac{r}{h}\cos\Phi\right]^{n\beta} \approx 1 + n\beta \cdot \frac{r}{h}\cos\Phi + \frac{1}{2}n\beta(n\beta - 1) \cdot \left(\frac{r}{h}\right)^2 \cos^2\Phi. \tag{24}$$

Here the time varying phase, $\Phi = 2\pi Nt$, where $N$ [rps] is rotational speed.

For both categories of airfoil self-noise and inflow-turbulence noise, the A-weighted sound power due to a blade segment of unit length $l_o = 1m$ takes the form (Eqs.19,22),

$$w(r,t) = q \cdot \left(\frac{v(l)}{v_o}\frac{r}{l}\right)^m \left(\frac{V(h)}{v_o}\right)^n F(t), \quad (25)$$

where $q$, as well as, $m$ and $n$ denote the free parameters that could be found from the $L_{WA}$, $v(l)$, and $V(h)$ measurements (see below). Mindful of the time period, $T = 1/N$, the average value of $w(r,t)$ (Eqs.24,25) becomes,

$$\langle w \rangle_r = q \cdot \left(\frac{v(l)}{v_o}\frac{r}{l}\right)^m \left(\frac{V(h)}{v_o}\right)^n \langle F \rangle_r \quad (26)$$

with the factor

$$\langle F \rangle_r = 1 + \frac{1}{4} n\beta(n\beta - 1)\left(\frac{l}{h}\right)^2, \quad (27)$$

which describes the influence of wind shear on sound power.

## 4. FULLY ENSONIFIED ZONE

Suppose the horizontal distance between the turbine and receiver meets the conditions, $\rho(\alpha) < \rho_1(\alpha)$, where

$$\rho_1(\alpha) = \mu \sqrt{\frac{(h-l)h}{\cos\alpha}} \quad \text{for} \quad -\frac{\pi}{2} < \alpha < +\frac{\pi}{2} \text{ (down receiver - Fig.2)}, \quad (28)$$

and

$$\rho_1(\alpha) = \infty \quad \text{for} \quad +\frac{\pi}{2} < \alpha < -\frac{\pi}{2} \text{ (upwind receiver - Fig.2)}. \quad (29)$$

For a such location, at every moment the entire blade of length $l$ (Fig.5) participates in the noise at the receiver. In other words, the sound rays from all points of the rotor plane reach the receiver.

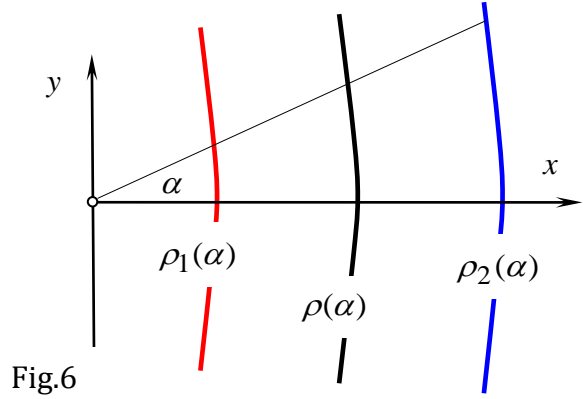

Fig.6

One can say that the receiver is fully insonified. To find the A-weighted sound power of the entire rotor plane, we calculate first the A-weighted sound power of a single blade by integration over its length $l$,

$$W_A = \int_0^l \langle w \rangle_r \, dr. \qquad (30)$$

Mindful of Eqs.(26,27) one gets,

$$W_A = \frac{q \cdot l}{m+1} \cdot \left(\frac{v(l)}{v_o}\right)^m \left(\frac{V(h)}{v_o}\right)^n \cdot \left[1 + \frac{1}{4} n\beta(n\beta - 1)\left(\frac{l}{h}\right)^2\right]. \qquad (31)$$

The above expression describes the effect of wind shear $\beta$ on wind turbine sound power: for $0 < n\beta < 0.5$ the value of $W_A$ decreases with $\beta$. Then, for $n\beta > 0.5$ the value of $W_A$ grows. The electric wind turbine power exhibits similar behavior [16].

When this wind shear effect is neglected, Eq.(31) simplifies to,

$$W_A \approx \frac{q \cdot l}{m+1} \cdot \left(\frac{v(l)}{v_o}\right)^m \left(\frac{V(h)}{v_o}\right)^n. \qquad (32)$$

Then we take into account three blades and substitute $3W_A$ into the definition of the A-weighted sound power level (Eq.16). Finally we arrive at the semi-empirical equation,

$$\boxed{L_{WA} \approx 10m \cdot \log \frac{v(l)}{v_o} + 10n \cdot \log \frac{V(h)}{v_o} + C}. \qquad (33)$$

The constant $C$, as well the parameters $m$ and $n$, can be determined from $L_{WA}$, $v(l)$ and $V(h)$ measurements. The constant $C$ depends on $m$ and the blade length, $l$,

$$C = 10 \log \frac{3 q l_o}{(m+1)W_o} + 10 \log \frac{l}{l_o}, \qquad l_o = 1m, \quad W_o = 10^{-12} W. \qquad (34)$$

The measurements reported in Refs. [10-15] confirm the dependence of $L_{WA}$ (Eq.33) on the blade tip speed, $v(l)$ (Eq.13), and on the wind speed at the hub height, $V(h)$ (Eq.20). Moreover, results by Rogers at all [17] indicate the increase of $L_{WA}$ with the logarithm of the blade length, $l$ (Eqs.33,34).

The engineering prediction models of the time average sound level, $L_{AeqT}$, are based on non-directional sound divergence over flat ground surface and air absorption (e.g. Refs.[18-23]). In reality the hub height $h$ and the horizontal distance to the receiver $\rho$ meet the conditions: $h < 100\,\text{m}$ and $\rho > 150\,\text{m}$. Thus, within 1 dB error, for the fully insonified zone (Eqs.28,29) one can write:

$$L_{AeqT} \approx L_{WA} - 10\log\left(\frac{2\pi\rho^2}{l_o^2}\right) - 0.005 \cdot \frac{\rho}{l_o} \,. \tag{35}$$

with $L_{WA}$ calculated from expression (33).

## 5. PARTIALLY INSONIFIED ZONE

The red and blue lines on Fig.6 described by the functions,

$$\rho_1(\alpha) = \mu\sqrt{\frac{(h-l)h}{\cos\alpha}}, \qquad \rho_2(\alpha) = \mu\sqrt{\frac{(h+l)h}{\cos\alpha}}, \qquad -(\pi/2) < \alpha < +(\pi/2), \tag{36}$$

with $\mu$ calculated from Eq.(11), determine the borders of the partially insonified zone. The sound energy inside this zone (Eq.36, Fig.6),

$$\rho_1(\alpha) < \rho(\alpha) = \mu\sqrt{\frac{hz}{\cos\alpha}} < \rho_2(\alpha), \tag{37}$$

comes form the red segment of the turbine blade (Fig.7). The B-C blue segment of the blade is not active. Therefore, the lower part of the rotor plane (below the blue horizontal line and above the blue arc) does not participate in noise immission at the receiver within the partially ensonified zone.

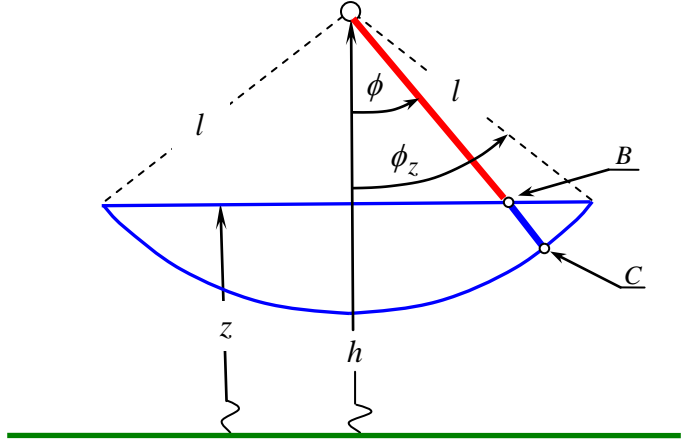
Fig.7

Note that height of the blue horizontal line is a function the receiver location (Eqs.36,37): close to the fully insonified zone, $\rho \to \rho_1$ (Fig.6), the blue line of the rotor plane tends to the lower tip of the plane, $z \to h-l$. On the other hand, when the receiver approaches the shadow zone, $\rho \to \rho_2$ (Fig.6), the blue horizontal line on Fig.7 rises up to $z \to h+l$.

In order to estimate the sound energy loss, $\Delta W_A(t)$, due to the non-active blade segment, we disregard the effect of wind shear on sound generation, $F(t)=1$ (Eqs.24,25) and integrate along the blue blade segment from $(h-z)/\cos\phi$ (point B) to $l$ (point C),

$$\Delta W_A(t) = \int_{(h-z)/\cos\phi}^{l} w(r)dr, \qquad \phi = 2\pi N t. \qquad (38)$$

Here

$$w(r) = q \cdot \left(\frac{v(l)}{v_o}\frac{r}{l}\right)^m \left(\frac{V(h)}{v_o}\right)^n, \qquad (39)$$

represents the A-weighted sound power from the blade segment of unit length $l_o = 1m$ (Eq.25). Integration of Eq.(38) brings about the time varying sound power

$$\Delta W_A(t) = W_A \cdot G(z,t), \qquad (40)$$

where $z$ denotes the height of the blue line – the boundary between the active and non-active part of the rotor plane (Fig.7), $W_A$ expresses the A-weighted sound power of an entire single blade of length $l$ (Eq.32), and

$$G(z,t) = 1 - \left[\frac{h-z}{l\cos\phi}\right]^{m+1}, \qquad \phi = 2\pi N t, \tag{41}$$

represents the sound power modulation factor. Finally, the A-weighted sound power of the active part of the rotor plane becomes (Eq.40),

$$W_A(z) = W_A \cdot [1 - \langle G \rangle]. \tag{42}$$

For the time period, $T = 1/N$ (Eq.14), the time average value of $G(t)$ becomes,

$$\langle G \rangle = \frac{1}{T}\int_0^T G(z,t)dt = \frac{\phi_z}{\pi} - \frac{1}{\pi}\int_0^{\phi_z}\left[\frac{\cos\phi_z}{\cos\phi}\right]^{m+1} d\phi. \tag{43}$$

From Figs.(6,7) and Eqs.(36,37) it follows that,

$$\cos\phi_z = \frac{\rho_1^2 + \rho_2^2 - 2\rho^2}{\rho_2^2 - \rho_1^2}. \tag{44}$$

When the receiver is close to the fully insonified zone, $\rho \to \rho_1$ (Eq.36), then $\phi_z \to 0$ (Eq.44) and the blue line – the boundary between the active and non-active part of the rotor plane gravitates to the lower tip of the rotor plane, $z \to h-l$ (Fig.7). In this case the entire rotor plane takes part in a sound power emission to the receiver. It is in agreement with the above formulae: $\langle G \rangle \to 0$ follows $\phi_z \to 0$ and then the A-weighted sound power for the partly insonified zone, $W_A(h-l)$ (Eq.42), becomes equal to the A-weighted sound power for the fully insonified zone: $W_A$ (Eq.32).

Further away from the turbine and close to the shadow zone, $\rho \to \rho_2$ (Eq.36), formula (44) gives $\phi_z \to \pi$. This means that the boundary between the active and non-active part of the rotor plane (blue line on Fig.7) rises up to the upper tip of the rotor plane, $z \to h+l$. In this case Eq.(43) gives $\langle G \rangle \to 1$ and consequently the A-weighted sound power for the partly insonified zone vanishes: $W_A(h+l) = 0$. Note that for $\rho > \rho_2$ the receiver falls into the shadow zone without any turbine sound (Fig.6).

In Sect.3 it was mentioned that the measurements of the A-weighted sound power $L_{WA}$, and rotational speed, $N$, give the values of $4 < m < 7$ [10-12]. Setting $m = 5$ in Eq.(43) one gets,

$$\langle G \rangle = \frac{1}{\pi}\left\{\phi_z - \frac{1}{5}\sin\phi_z\cos\phi_z - \frac{4}{15}\sin\phi_z\cos^3\phi_z - \frac{8}{15}\sin\phi_z\cos^5\phi_z\right\}. \tag{45}$$

To find the dependence of $\langle G \rangle$ on the distance $\rho$ within the partly ensonified zone, we apply $\cos\phi_z$ (44) and

$$\sin\phi_z = 2\frac{\sqrt{\rho^2(\rho_2^2 + \rho_1^2) - \rho_1^2\rho_2^2 - \rho^4}}{\rho_2^2 - \rho_1^2}, \qquad (46)$$

where distances $\rho_1$ and $\rho_2$ are defined by Eq.(36) and shown on Fig.6. By applying Eqs.(42,44,45,46) one finds the A-weighted sound power,

$$W_A(\rho) = W_A[1 - \langle G \rangle], \qquad (47)$$

where

$$\langle G \rangle = \frac{1}{\pi}\left\{\arccos\frac{2\rho^2 - \rho_1^2 - \rho_2^2}{\rho_2^2 - \rho_1^2} - \frac{2}{5}\frac{(\rho_1^2 + \rho_2^2 - 2\rho^2)\sqrt{\rho^2(\rho_1^2 + \rho_2^2) - \rho_1^2\rho_2^2 - \rho^4}}{(\rho_2^2 - \rho_1^2)^2} - \ldots\right\} \qquad (48)$$

Substituting $\rho = \rho_1$ and $\rho = \rho_2$ one arrives at $\langle G \rangle_1 = 0$ and $\langle G \rangle_2 = 1$, respectively. Finally, for the receiver within the partially insonified zone, $\rho_1 < \rho < \rho_2$ (Eq.37), the formula below

$$\boxed{L_{AeqT} \approx L_{WA} + \Delta L_{WA}(\rho) - 10\log\left(\frac{2\pi\rho^2}{l_o^2}\right) - 0.005 \cdot \frac{\rho}{l_o}}, \qquad (49)$$

gives the time average sound level. Here $L_{WA}$ is determined by the semi-empirical equation (33) and the correction to the A-weighted sound power level equals,

$$\boxed{\Delta L_{WA}(\rho) = 10\log[1 - \langle G \rangle]}, \qquad (50)$$

with $\langle G \rangle$ calculated from the integral (43).

**Example 2**

For wind conditions as in Example 1, the transition zone is determined by $\rho_1 = 670m$ and $\rho_2 = 1250m$. The correction $\Delta L_{WA}(\rho)$, for the partially ensonified zone $\rho_1 < \rho < \rho_2$ (Eq.37) is plotted on Fig. 8.

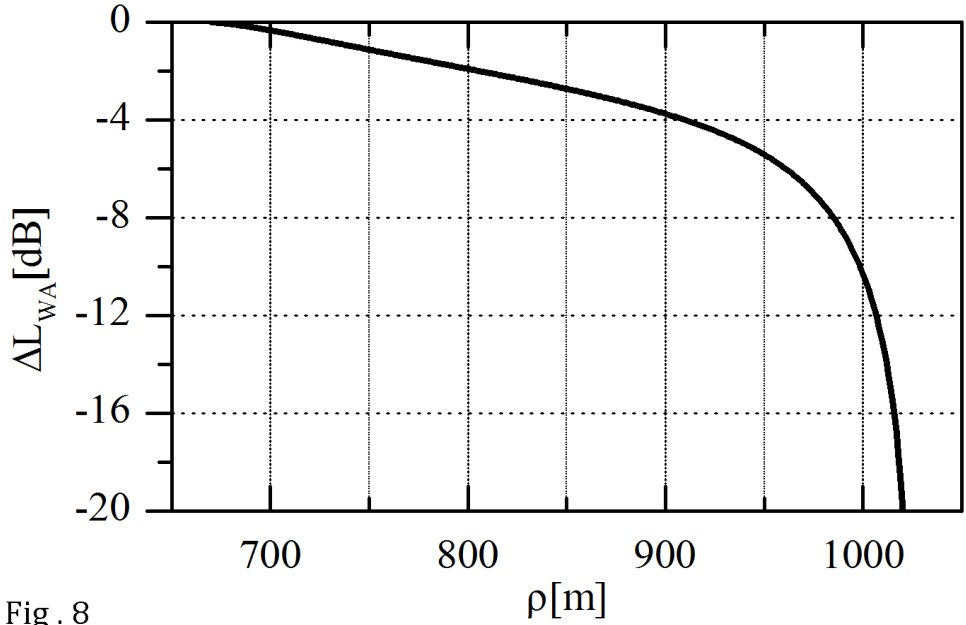
Fig. 8

### 5. CONCLUSIONS

The semi-empirical theory presented here is based on measurements of : $L_{WA}$ - A-weighted sound power level, $v(l)$ - blade tip speed, and $V(h)$ - wind speed at the hub height. Due to wind shear induced refraction the turbine blade segment do-, and do not contribute to the sound energy at the receiver (Fig.7). To find the A-weighted power of sound emitted from the sound active segment, the concept of the blade of unite length, $l_o$, was introduced (Eq.26).

When the receiver is located within the fully ensonifeid zone (Eqs.28,29 and Fig.6), formulae (33) and (35) can be used for calculation of the time-average sound level, $L_{AeqT}$. Further away from the turbine, within the partially ensonified zone (Eqs.36,37 and Fig.6), expressions (49) and (50) can be applied. Measurements indicate (Fig.2) that at the same horizontal distance from the turbine $\rho$, the downwind value of $L_{AeqT}^{(down)}$ exceeds the upwind value of $L_{AeqT}^{(up)}$. To quantify this effect we make use of Eqs.(49) and (50),

$$L_{AeqT}^{(down)} - L_{AeqT}^{(up)} = -10\log\left[1-\langle G\rangle_\rho\right]. \tag{51}$$

For $\rho = 1000\,m$ graph on Fig.8 gives $\Delta L_{WA} = 10\,dB$ : more or less the theory complies with the experiment (Fig.2).

The method presented here cannot be applied in hilly terrain or in the coastal area.

**FIGURE CAPTIONS**

Fig.1. Down- and up-wind noise propagation of wind turbine noise with the hub height $h$ and rotor diameter $l$

Fig.2. Time variations of the A-weighted sound pressure level at the horizontal distance $\rho = 1000$ m, for down- and upwind propagation (Ref.2)

Fig.3. The rays from a low- and high tips of the rotor plane graze the ground at the horizontal distances $\rho_1(\alpha)$ and $\rho_2(\alpha)$, respectively (Eq.36).

Fig.4. Horizontal distance to the shadow zone $\rho(\alpha)$ (Eq.37). The wind blows at the speed $V$ along the *x*-axis.

Fig.5. The blade segment of the unite length, $l_o = 1m$, rotates with the speed $v(r) = 2\pi Nr$ at the distance $0 < r < l$ from the hub.

Fig.6. Boundaries of fully- and partly ensonified zones: $\rho_1(\alpha)$ and $\rho_2(\alpha)$ (Eqs.11,36).

Fig.7. Sound active (red line) and not-active (BC blue line) segments of the blade.

Fig.8. The effect of refraction within the partly ensonified zone $\rho_1 = 670m$ and $\rho_2 = 1250m$ (Eqs.36,37 and Fig.6): correction $\Delta L_{WA}$ (Eqs.43,44,50) to the A-weighted sound power level, $L_{WA}$ (Eq.33), as a function of the horizontal distance $670 < \rho < 1250 \, \text{m}$.